\documentstyle[11pt]{article}

\newcommand{\sect}[1]{\setcounter{equation}{0}\section{#1}}
\newcommand{\subsect}[1]{\subsection{#1}}

\renewcommand{\theequation}{\arabic{section}.\arabic{equation}}

\def\be{\begin{equation}}
\def\ee{\end{equation}}
\def\bea{\begin{eqnarray}}
\def\eea{\end{eqnarray}}

\def\1{\'{\i}}                           
\def\R{{\rm I\kern-.2em R}}

\def\ap{A_+}
\def\am{A_-}
\def\aa{A}
\def\bb{M}

\def\jp{J_+}
\def\jm{J_-}
\def\jj{J_3}

\def\ctea{\omega}
\def\para{\omega^2}
\def\deforn{\lambda}
\def\qq{Q}
\def\kk{K}
\def\fun{\gamma}
\def\s{{\sigma}}

\def\co{\Delta}

\def\w{\omega}

\def\s{{\sigma}} 

\def\>#1{{\bf #1}}                 
\def\1{\'{\i}}                           
\def\R{\rm I\kern-.2em R} 
\def\C{\rm I\kern-.5em C}

\def\pois#1#2{\left\{ {#1},{#2} \right\}}

\begin{document}
 
\thispagestyle{empty}
\hfill\today 
\vspace{4.5cm}

\begin{center} {\LARGE{\bf{Long range integrable oscillator chains}}} 

{\LARGE{\bf{from quantum algebras}}} 

\vspace{2.4cm} 

ANGEL BALLESTEROS and FRANCISCO J. HERRANZ\vspace{.2cm}\\
{\it Departamento de F\1sica, Universidad de Burgos\\
Pza. Misael Ba\~nuelos s.n., 09001-Burgos, Spain} 
\vspace{.3cm}

\end{center} 
  
\bigskip

\begin{abstract} 

Completely integrable Hamiltonians   defining classical mechanical
systems of $N$ coupled oscillators are obtained from Poisson realizations
of Heisenberg--Weyl, harmonic oscillator and $sl(2,\R)$ coalgebras. Various
completely integrable deformations of such systems are constructed by
considering quantum deformations of these algebras. Explicit expressions
for all the deformed Hamiltonians and constants of motion are given,
and the long-range nature of the interactions is shown to be linked
to the underlying coalgebra structure. The relationship between 
oscillator systems induced from the $sl(2,\R)$ coalgebra and angular
momentum chains is presented, and a non-standard integrable deformation of
the hyperbolic Gaudin system is obtained.

\end{abstract} 

\bigskip

\vfill\eject


\sect{Introduction}

The construction of integrable systems is an outstanding application of Lie
algebras in both classical and quantum mechanics \cite{OP,Per}. In fact,
the very definition of integrability is based on the concept of
involutivity of the conserved quantities with respect to a (either Poisson
or commutator) Lie bracket. During last years, many new
results concerning deformations of Lie algebras and groups have been
obtained \cite{CP}, and the question concerning the connection between
these new algebraic structures and integrability properties arises as a
keystone for future developments in the subject. From our point of view,
such analysis should be based on the basic role that the coalgebra
structure (i.e., the existence of an homomorphism $\Delta$ between the
algebra $A$ and $A\otimes A$, called the coproduct) plays in quantum
deformations.

In this paper, we follow this idea and deal with a general and systematic
construction of integrable systems from coalgebras that has been
introduced very recently \cite{BCR,BR}. Within it, the coproduct map plays
an essential role: it propagates integrability from the one-particle
Hamiltonian (which is defined as a certain realization of a given
function of the algebra generators) to the $N$-particle one. Several
classical realizations of the Gaudin system have been presented as the
first examples of this coalgebra-induced construction. In this
framework, quantum algebras (which are just coalgebra deformations) can be
interpreted as dynamical symmetries that generate in a direct way a large class
of integrable deformations. This has been the case for the
Gaudin--Calogero systems \cite{Cal} with respect to the (standard) quantum
deformation $U_z(sl(2,\R))$, that can be used to derive explicitly a family
of completely integrable Hamiltonians that reduce to the Gaudin--Calogero
one in the ``classical" limit $z\to 0$ \cite{BCR}.

In the same way, a sort of
Ruijsenaars--Schneider Hamiltonian was derived from a quantum deformation
of (1+1) Poincar\'e algebra \cite{BRWig}, and also a simple example of
oscillator chain was obtained from the non-deformed oscillator algebra
$h_4$ \cite{BR}. However, in order to get a deeper insight into this
general method and its consequences, it seems appropriate to construct
some new models. This is the aim of the present paper, in which we shall
concentrate on the explicit construction and analysis of classical
mechanical oscillator chains obtained from (deformed and non-deformed)
Heisenberg--Weyl $h_3$, harmonic oscillator $h_4$ and $sl(2,\R)$
coalgebras, all of them realized in terms of functions on the classical
phase space endowed with the canonical Poisson bracket. We recall that 
quantum  algebra deformations of $h_3$ and
$h_4$ have been fully classified in \cite{Heis,Osc}, and results concerning
$sl(2,\R)$ are basic in quantum algebra theory and can be found,
for instance, in \cite{CP}.

In the next section we briefly recall the general construction of
\cite{BR} and fix the notation. It is important to stress that the
complete integrability of the resulting Hamiltonians by no means depends
on the explicit form of the coproduct map, that can be either deformed or
classical. Therefore, in the following sections we will pay attention to
both types of situations, and we shall emphasize the fact that the
underlying algebraic structure provides explicit
expressions for the constants of motion in a straightforward way.
However, we shall not address here the study of the dynamical contents of
the (to our knowledge) new oscillator chains that will be introduced.

The simplest coalgebra structures are provided by the Heisenberg--Weyl
algebra, and are analysed in section 3. In the non-deformed case, the
coalgebra leads to an oscillator chain whose integrability cannot be
immediately proven due to the ``degeneracy" of the Heisenberg phase space
realization. On the contrary, a two-parameter quantum deformation
of $h_3$ \cite{Heis} gives rise in a straightforward way to the
corresponding integrable deformation, including the explicit form for the
constants of motion. Therefore, degeneracies of
the integrals of motion appearing in the non-deformed model are removed.

The same procedure is carried out for the oscillator algebra $h_4$ in
section 4. The Hamiltonian there induced from the non-deformed coalgebra is
the same as the one derived in the Heisenberg case, but now its
complete integrability is easily demonstrated by the existence of an extra
Casimir function (therefore, the degeneracy problem can be circumvented 
by taking into account that $h_3 \subset h_4$). The non-standard quantum
deformation of $h_4$ \cite{Osc} is also realized in terms of canonical
coordinates and gives rise to another integrable chain.

Section 5 deals with oscillator chains obtained from $sl(2,\R)$ coalgebras.
In contrast with the previous cases, the non-deformed $sl(2,\R)$
structure provides a chain of non-interacting oscillators through a
linear  Hamiltonian of the type $\jp + \alpha\,\jm$. In this way, the
inclusion of the  non-standard  deformation of $sl(2,\R)$ \cite{Ohn}  can
be interpreted as a direct algebraic implementation of a certain type of
long-range interaction involving the momenta. In this section we also
present the deep relationship between these $sl(2,\R)$ oscillator chains
and classical spin models, and we explicitly construct the non-standard
deformation of the hyperbolic Gaudin system \cite{Gau}. As it happened
with the standard deformation \cite{BR}, the non-standard one is shown to
generate a sort of variable range exchange \cite{Inoz}. The construction of
(deformed and non-deformed) anharmonic chains is also studied, thus
showing the  number of integrable systems that can be easily derived by
following the present approach.


\section{From coalgebras to integrable Hamiltonians}

The main result of \cite{BR} can be summarized as follows: any
coalgebra $(A,\co)$ with Casimir element $C$ can be considered as the
generating symmetry that, after choosing a non-trivial representation,
gives rise in a   systematic way to a large family of integrable
systems. We shall consider here classical mechanical systems only and,
consequently, we shall make use of Poisson realizations $D$ of Lie and
quantum algebras of the form $D:A\rightarrow C^\infty (q,p)$. However, we
recall that the formalism is also directly applicable to quantum
mechanical systems.

Let
$(A,\Delta)$ be a (Poisson) coalgebra with generators $X_i$
$(i=1,\dots,l)$ and Casimir element $C(X_1,\dots,X_l)$. Therefore, the
coproduct $\Delta:A\rightarrow A\otimes A$ is a Poisson map. Let us
consider the $N$-th coproduct $\co^{(N)}(X_i)$ of the generators \be
\co^{(N)}:A\rightarrow A\otimes A\otimes \dots^{N)}\otimes A 
\ee
which is obtained (see \cite{BR}) by applying recursively the two-coproduct
$\co^{(2)}\equiv \co$ in the form
\be
\co^{(N)}:=(id\otimes id\otimes\dots^{N-2)}\otimes id\otimes
\co^{(2)})\circ\co^{(N-1)}.
\label{fl}
\ee
 By taking into account that the $m$-th
coproduct $(m\leq N)$ of the Casimir $\co^{(m)}(C)$ can be embedded into
the tensor product of $N$ copies of $A$ as
\be
\co^{(m)}:A\rightarrow \{A\otimes A\otimes \dots^{m)}\otimes A\} \otimes
\{1 \otimes 1\otimes \dots^{N-m)}\otimes 1\},
\ee
it can be shown that, \be
\pois{\co^{(m)}(C)}{\co^{(N)}(X_i)}_{A\otimes
A\otimes\dots^{N)}\otimes A}=0 \qquad
i=1,\dots,l \quad 1\leq m\leq N .
\label{za}
\ee

With this in mind it can be proven \cite{BR} that, if ${\cal
H}$ is an {\it arbitrary} (smooth) function of the
generators of $A$, the $N$-particle Hamiltonian defined on $A\otimes
A\otimes\dots^{N)}\otimes A$ as the $N$-th coproduct of ${\cal H}$
\be
H^{(N)}:=\co^{(N)}({\cal{H}}(X_1,\dots,X_l))=
{\cal{H}}(\co^{(N)}(X_1),\dots,\co^{(N)}(X_l)),
\label{htotg}
\ee
fulfils
\be
\pois{C^{(m)}}{H^{(N)}}_{A\otimes
A\otimes\dots^{N)}\otimes A}=0 \qquad 1\leq m\leq N 
\label{za1}
\ee
where the $N$ functions $C^{(m)}$ ($m=1,\dots,N$)
are defined through the coproducts of the Casimir $C$
\be
C^{(m)}:= \co^{(m)}(C(X_1,\dots,X_l))=
C(\co^{(m)}(X_1),\dots,\co^{(m)}(X_l))
\label{Ctotg}
\ee
and all the integrals of motion $C^{(m)}$ are in involution
\be
\pois{C^{(m)}}{C^{(n)}}=0 \qquad \forall\,m,n=1,\dots,N.
\label{cor3}
\ee

Therefore, provided a realization of $A$ on a one-particle phase space is
given, the
$N$-particle Hamiltonian $H^{(N)}$ will be a function of $N$ canonical
pairs $(q_i,p_i)$ and is, by construction, completely  integrable with
respect to the ordinary Poisson bracket 
\be
\{f,g\}=\sum_{i=1}^N\left(\frac{\partial f}{\partial q_i}
\frac{\partial g}{\partial p_i}
-\frac{\partial g}{\partial q_i} 
\frac{\partial f}{\partial p_i}\right) .
\label{poisbra}
\ee
Moreover, its
constants of motion will be given by the $C^{(m)}$ functions, all of
them functionally independent since each of them depends on the first $m$
pairs $(q_i,p_i)$ of canonical coordinates. Note that with such
one-particle realizations the first Casimir $C^{(1)}$ will be a number,
and we are left with $N-1$ constants of motion with respect to $H^{(N)}$.

In particular, this result can be applied to universal enveloping algebras
of Lie algebras $U(g)$ \cite{BCR}, since they are always endowed with a
natural (primitive) Hopf algebra structure of the form
$\co^{(2)}(X_i)= X_i\otimes 1+1\otimes X_i $, being $X_i$  any generator 
of $g$. Moreover, since quantum algebras are also (deformed) coalgebras
$(A_z,\Delta_z)$, any function of the generators of a given quantum
algebra with Casimir element $C_z$ will provide, under a chosen deformed
representation, a completely integrable Hamiltonian.


\sect{Oscillator chains from Heisenberg--Weyl coalgebras}

The obtention of integrable oscillator chains by using the previous
approach can be achieved by selecting Poisson coalgebras $(A,\co)$ such
that the one-dimensional harmonic oscillator Hamiltonian with angular
frecuency $\w$ (and unit mass) could be written as the phase space
representation $D$ of a certain function ${\cal H}$ of the generators of
$A$: 
\be
H=D ({\cal H})=p^2 + \w^2 q^2  .
\label{oh}
\ee
It seems natural to consider the non-deformed Heisenberg--Weyl coalgebra
(whose generators are essentially the canonical coordinates and a central
element) as the first ``dynamical coalgebra" for oscillator chains.


\subsect{Non-deformed $U(h_3)$ coalgebra}

The $h_3$ Lie--Poisson algebra
\be
\{\am,\ap\}=\bb\qquad \{\am,\bb\}=0\qquad 
\{\ap,\bb\}=0 
\label{bb}
\ee
is endowed with a Poisson coalgebra structure by means of the usual
primitive coproduct
\be
\Delta(X)= X \otimes 1 +1\otimes X  
\qquad X\in \{\am,\ap,\bb\} 
\label{ba}
\ee
which is a Poisson algebra homomorphism between $h_3$ and $h_3\otimes
h_3$. The Casimir element ${\cal C}$ for this algebra is the central
generator $\bb$.

A natural one-particle phase space realization $D$ of $h_3$ is
given by:
\be
f_-^{(1)}=D(\am)=\ctea_1  q_1\qquad 
f_+^{(1)}=D(\ap)=p_1
\qquad
f_M^{(1)}= D(\bb)=\ctea_1 .
\label{bd}
\ee
Obviously, if we consider the quadratic function ${\cal H}$ of the $h_3$
generators \be
{\cal H}= \ap^2+ \am^2,
\label{fun}
\ee
the associated one-particle Hamiltonian will be
\be
H^{(1)}=D({\cal H})=(f_+^{(1)})^2+(f_-^{(1)})^2= p_1^2 + \ctea_1^2  
q_1^2,
\label{be}
\ee
and the Casimir function is just the angular frecuency
\be
C^{(1)}=D({\cal C})=\ctea_1.
\label{cm}
\ee

Now, the coproduct map (\ref{ba}) will give us, under a $D\otimes D$
realization, three two-particle phase space functions that close again
$h_3$ under the usual Poisson bracket $\pois{q_i}{p_j}=\delta_{i\,j}$:
\bea
&&f_-^{(2)}=(D\otimes D)(\Delta(\am))=\ctea_1 q_1+\ctea_2  q_2\cr
&&f_+^{(2)}=(D\otimes D)(\Delta(\ap))= p_1 + p_2\cr
&&f_M^{(2)}=(D\otimes D)(\Delta(\bb))=\ctea_1+\ctea_2 .
\label{bf}
\eea
By folowing (\ref{htotg}), the two-particle Hamiltonian $H^{(2)}$ will be
given by the realization of the coproduct of ${\cal H}$ and reads
\bea
&&H^{(2)}=(D\otimes D)(\Delta({\cal H}))= (f_+^{(2)})^2+(f_-^{(2)})^2\cr
&&\qquad
=p_1^2+p_2^2+ \ctea_1^2  q_1^2+\ctea_2^2  q_2^2 +
2(p_1 p_2 +\ctea_1\ctea_2   q_1 q_2),
\label{bg}
\eea
which is just the Hamiltonian defining a pair of coupled oscillators with
frecuencies $\ctea_1$ and $\ctea_2$ and whose interaction depends on the
momenta. The  constant of motion $C^{(2)}$
would be given by (\ref{Ctotg}), i.e. the representation of the coproduct
of the Casimir
\be C^{(2)}=(D\otimes D)(\Delta({\cal C}))=\ctea_1+\ctea_2.
\label{bh}
\ee
In this case, since the Casimir is a single generator whose
representation is a real constant, its coproduct is just the sum of
he two frecuencies.

The method sketched in section 2 allows us to generalize this result to
the $N$-dimensional case in a straightforward way. Since, by using
(\ref{fl}), the $m$-th coproduct of a primitive generator reads
\bea
&&\!\!\!\!\!\!\!\!\co^{(m)}(X_i)=X_i\otimes 1\otimes 1\otimes
\dots^{m-1)}\otimes 1 \cr
&&\qquad\qquad\quad + 1\otimes X_i\otimes 1\otimes\dots^{m-2)}\otimes
 1 +
\dots \cr
&&\qquad\qquad\qquad\quad + 1\otimes
1\otimes\dots^{m-1)}\otimes 1\otimes X_i,
\label{fo}
\eea
the $m$-dimensional particle phase space realization of $(h_3,\co)$ will
be:
\bea
&&f_-^{(m)}=(D\otimes\dots {}^{m)}\otimes D)(\Delta^{(m)}(\am))=
\sum_{i=1}^m \ctea_i   q_i\cr
&&f_+^{(m)}=(D\otimes\dots {}^{m)}\otimes D)(\Delta^{(m)}(\ap))
=\sum_{i=1}^m p_i\cr
&&f_M^{(m)}=(D\otimes\dots {}^{m)}\otimes D)(\Delta^{(m)} (\bb))=
\sum_{i=1}^m \ctea_i.
\label{bi}
\eea
Note that these three functions generate again $h_3$. Therefore, the
corresponding $N$-dimensional Hamiltonian is: \bea
&&H^{(N)}=(D\otimes\dots {}^{N)}\otimes D)(\Delta^{(N)} ({\cal H}))
=   (f_+^{(N)})^2+  (f_-^{(N)})^2\cr
&&\qquad\quad 
=\sum_{i=1}^N( p_i^2 +\ctea_i^2   q_i^2)  + 2 \sum_{i<j}^N (p_i p_j +
\ctea_i \ctea_j   q_i q_j).
\label{bj}
\eea
In the same manner, the $N$ constants of motion provided by the
formalism would be
\be C^{(m)}=(D\otimes\dots {}^{m)}\otimes
D)(\Delta^{(m)} ({\cal C})) =f_M^{(m)}=\sum_{i=1}^m \ctea_i
\qquad m=1,\dots,N.
\label{bk}
\ee

Apparently, we could conclude that in this case the formalism provides
trivial constants of motion (\ref{bk}) only. This problem is due to the
extreme simplicity of the chosen algebra and representation, and the
Hamiltonian (\ref{bj}) will be proven to be completely integrable in
section 4 by making use of the embedding of $h_3$ into the harmonic
oscillator algebra $h_4$, which is the natural framework to construct this
kind of non-deformed chains.  However, the $h_3$ case provides an
interesting benchmark for the use of quantum algebras to obtain integrable
deformations. As we shall see in the sequel, the deformation will provide
non-trivial Casimirs, thus breaking the degeneracy in the representation
and turning the analogues of (\ref{bk}) into non-trivial functions.


\subsect{Two-parameter quantum deformation $U_{z,\deforn}(h_3)$}

Quantum deformations of the Heisenberg--Weyl algebra have been fully
classified in \cite{Heis}. We shall now make use of the
two-parameter quantum algebra $U_{z,\deforn}(h_3)$ with deformed coproduct
\bea
&&\Delta(\am)=  \am \otimes 1+ 1\otimes \am \qquad
\Delta(\bb)=\bb \otimes e^{z \am} + e^{-z \am} \otimes \bb   \cr
&&\Delta(\ap)=\ap \otimes e^{z \am} + e^{-z \am} \otimes \ap   
+ \deforn \bb \otimes \am e^{z \am},
\label{ca}
\eea
which is compatible with the non-deformed Poisson
brakets \be
\{\am,\ap\}=\bb\qquad \{\am,\bb\}=0\qquad 
\{\ap,\bb\}=0  .
\label{cb}
\ee
As a consequence, the Casimir coincides again with the $M$ generator
and the one-particle representation $D$ (\ref{bd}) is also valid for
the deformed case. 

In fact, the deformation arises with the coproduct and, consequently,
it will become apparent in the two-particle system. Although the
phase space realizations on each space are classical ones, their
composition law (\ref{ca})  is not, and leads to the following functions
\bea
&&f_-^{(2)}=(D\otimes D)(\Delta(\am))={\ctea_1 q_1}+{\ctea_2 q_2}\cr
&&f_+^{(2)}=(D\otimes D)(\Delta(\ap))=(p_1+   \deforn  {\ctea_1
\ctea_2 q_2}) e^{z {\ctea_2 q_2}} +  p_2 e^{- z {\ctea_1 q_1}}\cr
&&f_M^{(2)}=(D\otimes D)(\Delta(\bb))=\ctea_1 e^{z {\ctea_2 q_2}}+
\ctea_2 e^{-z{\ctea_1 q_1}},
\label{cf}
\eea
whose Poisson brackets will close again an $h_3$ algebra. If
the function ${\cal H}$ given in (\ref{fun}) is considered,
these expressions will lead to a deformed two-particle Hamiltonian
\bea
&&H_{z,\deforn}^{(2)}=(D\otimes D)(\Delta({\cal H}))= (f_+^{(2)})^2+
(f_-^{(2)})^2\cr && 
\qquad = (p_1 + \deforn \ctea_1 \ctea_2 q_2)^2 e^{2 z
\ctea_2 q_2}  +
p_2^2 e^{-2 z
\ctea_1 q_1}   + \ctea_1^2  q_1^2 + \ctea_2^2  q_2^2  \cr
&&\qquad \qquad 
+ 2 \left\{(p_1 + \deforn \ctea_1 \ctea_2 q_2) p_2 e^{-z
(\ctea_1 q_1-\ctea_2 q_2)} + \ctea_1  \ctea_2   q_1 q_2\right\},
\label{cg}
\eea
whose power series expansion around zero vales for the deformation
parameters is
\bea
&&H_{z,\deforn}^{(2)}=H^{(2)} + 2 z (p_1 + p_2) (\ctea_2  q_2 p_1-
\ctea_1 q_1  p_2) + 2 \deforn \ctea_1 \ctea_2 q_2 (p_1+p_2) \cr
&&\qquad\quad +
2z\deforn \ctea_1 \ctea_2 q_2 (\ctea_2   q_2 p_1-
\ctea_1  q_1 p_2 + \ctea_2 q_2 (p_1+p_2)) + \mbox{o} (z^2,\deforn^2) 
\label{exp}
\eea
where $H^{(2)}$ is the non-deformed Hamiltonian (\ref{bg}). 

Deformed Casimirs are now
\be
C_{z,\deforn}^{(1)}=D({\cal C})=\ctea_1\qquad
C_{z,\deforn}^{(2)}=(D\otimes D)(\Delta({\cal C}))=
\ctea_1 e^{z {\ctea_2 q_2}}+\ctea_2 e^{-z {\ctea_1 q_1}}.
\label{ch}
\ee
In particular, due to the deformed coproduct (\ref{ca}) for the central
generator $M$, the two-particle Casimir is no longer a constant, and the
degeneracy problem is solved.

The construction of the $N$-dimensional system is based onto the
general $m$-dimensional particle phase space realization given by:
\bea
&&f_-^{(m)}=(D\otimes\dots {}^{m)}\otimes D)(\Delta^{(m)}(\am))=
\sum_{i=1}^m \ctea_i  q_i\cr
&&f_+^{(m)}=(D\otimes\dots {}^{m)}\otimes D)(\Delta^{(m)}(\ap))
=\sum_{i=1}^m \pi_i^{(m)}  e^{z \qq_i^{(m)}(q)}\cr
&&f_M^{(m)}=(D\otimes\dots {}^{m)}\otimes D)(\Delta^{(m)} (\bb))=
\sum_{i=1}^m \ctea_i e^{z \qq_i^{(m)}(q)}
\label{ci}
\eea
where we define the auxiliary quantities
\be 
 \pi_i^{(m)}=p_i + \deforn \ctea_i \qq_{+,i}^{(m)}(q).
\label{cj}
\ee
The various $\qq$-functions, that we shall often use from
now on, are introduced in the appendix; within all of them, any sum 
defined on an empty set of indices will be assumed to be zero.

 The final expression
for the $N$-dimensional deformed chain is:
\bea
&&\!\!\!\!\! 
H_{z,\deforn}^{(N)}=(D\otimes\dots {}^{N)}\otimes
D)(\Delta^{(N)} ({\cal H})) =   (f_+^{(N)})^2+ (f_-^{(N)})^2\cr
&& \!\!\!\!\!
=\sum_{i=1}^N \left( (\pi_i^{(N)})^2 e^{2 z \qq_i^{(N)}(q)} 
+ \ctea_i^2  q_i^2\right) 
+ 2 \sum_{i<j}^N \left(\pi_i^{(N)}\pi_j^{(N)} e^{z \qq_{ij}^{(N)}(q)} +
\ctea_i \ctea_j   q_i q_j\right) .
\label{cmm}
\eea
As it was expected, the $N-1$ constants of motion depend on the canonical
coordinates $q_i$ and read $(m=2,\dots,N)$:
\be
C_{z,\deforn}^{(m)}=(D\otimes\dots {}^{m)}\otimes
D)(\Delta^{(m)} ({\cal C})) =f_M^{(m)}=
\sum_{i=1}^m \ctea_i e^{z \qq_i^{(m)}(q)}.
\label{cn}
\ee


\sect{Oscillator chains from harmonic
oscillator coalgebras}

As we shall see now, the harmonic oscillator algebra $h_4$ will provide
a natural setting for the construction of non-deformed chains. We shall
also analyse a new system obtained from the (uniparametric) non-standard
deformation of $h_4$ introduced in \cite{Osc}.


\subsect{Non-deformed $U(h_4)$ coalgebra}

We consider the Poisson oscillator algebra defined by
\be
\{\aa,\ap\}=\ap\qquad 
\{\aa,\am\}=-\am\qquad \{\am,\ap\}=\bb\qquad \{\bb,\,\cdot\,\}=0.
\label{db}
\ee
As usual, the primitive coproduct
\be
\Delta(X)= X \otimes 1 +1\otimes X  
\qquad X\in \{\aa,\am,\ap,\bb\} 
\label{da}
\ee
is compatible with (\ref{db}) and we have two Casimirs: $\bb$ and
\be
{\cal C}=\aa\bb - \am \ap.  
\label{dc}
\ee
The latter will play a relevant role in what follows. Obviously, $h_3$ is
a subalgebra of $h_4$, and a natural phase space realization for $h_4$ is
given by
\bea
&&f_-^{(1)}=D(\am)=\ctea_1  q_1\qquad 
f_+^{(1)}=D(\ap)=p_1\cr
&&f_M^{(1)}= D(\bb)=\ctea_1 \qquad f^{(1)}=D(\aa)=q_1 p_1   , 
\label{dd}
\eea
and it is characterized by the values of the Casimirs $D(\bb)=\ctea_1$ and
$C^{(1)}=D({\cal C})=0$.

By following our construction, we shall consider again the quadratic
function \be
{\cal H}= \ap^2+  \am^2, 
\label{dda}
\ee
that gives rise to the one dimensional harmonic oscillator through $D$:
\be
H^{(1)}=D({\cal H})= p_1^2 +\ctea_1^2   q_1^2.
\label{de}
\ee
Note that (\ref{dda}) and (\ref{de}) are just the same expressions we had
in the $h_3$ construction.

The coproduct (\ref{da}) induces the two-particle phase space realization:
\bea 
&& f_-^{(2)}=\ctea_1  q_1+\ctea_2 q_2\qquad
f_+^{(2)}=p_1 + p_2\cr
&& f_M^{(2)}=\ctea_1+\ctea_2\qquad
 f^{(2)}=q_1 p_1 + q_2 p_2 .
\label{df}
\eea 
From (\ref{df}) and (\ref{dda}) we are  just lead to the two-dimensional
Hamiltonian (\ref{bg}). But there exists a crucial difference with respect
to the $h_3$ construction: the $(D\otimes D)$ realization of the coproduct
of  ${\cal C}$ is no longer a constant: namely,
\be
C^{(2)}=(\ctea_1  p_2 - \ctea_2 p_1) (q_2 - q_1).
\label{dh}
\ee
Therefore, the complete integrability of (\ref{bg}) is easily proven
through the use of the new Casimir function ${\cal C}$ of the $h_4$
coalgebra.

Similarly, the construction of the $m$-dimensional particle phase space
realization in terms of $C^{(1)}=0$ representations is guided by
(\ref{fo}):
\be 
 f_-^{(m)}= 
\sum_{i=1}^m \ctea_i q_i\qquad
 f_+^{(m)}=\sum_{i=1}^m p_i\qquad
 f_M^{(m)}=
\sum_{i=1}^m \ctea_i \qquad
 f^{(m)} 
=\sum_{i=1}^m q_i p_i .
\label{di}
\ee
The $N$-dimensional particle integrable system linked to (\ref{dda}) is
clearly (\ref{bj}). Explicit integrability is thus provided by the
constants given through the different coproducts $(m=2,\dots,N)$ for the
Casimir (\ref{dc}), that can be explictly written as:
\be
C^{(m)} =
\sum_{i=1}^m q_i p_i \left\{ \sum_{j=1}^{i-1}\ctea_j +
\sum_{l=i+1}^{m}\ctea_l \right\}
- \sum_{i=1}^m p_i \left\{   \qq_{+,i}^{(m)}(q) +
\qq_{-,i}^{(m)}(q) \right\} .
\label{dk}
\ee


Its
Poisson coalgebra expressions are found to be

\subsect{Non-standard quantum deformation $U_z(h_4)$}

This quantum algebra, firstly introduced in \cite{Osc}, is characterized
by a deformation of both the coproduct and the commutation rules. The
former is given by
\bea
&&\Delta(\am)=  \am \otimes 1+ 1\otimes \am \qquad
\Delta(\bb)=\bb \otimes 1 + 1 \otimes \bb   \cr
&&\Delta(\ap)=\ap \otimes e^{z \am} + e^{-z \am} \otimes \ap   
+ 2 z \aa \otimes \bb e^{z \am} \cr
&&\Delta(\aa)=\aa \otimes e^{z \am} + e^{-z \am} \otimes \aa 
\label{ea}
\eea
and the latter are translated in Poisson terms as
\bea
&&\{\aa,\ap\}=\ap \cosh z\am - z\aa \bb e^{z \am}\qquad
\{\bb,\,\cdot\,\}=0\cr &&\{\aa,\am\}=-\frac {\sinh z\am}{z}\qquad
\{\am,\ap\}=\bb e^{z \am} .
\label{eb}
\eea
It is straightforward to check that (\ref{ea}) is a Poisson map with
respect to (\ref{eb}), and that $\bb$ and \be
{\cal C}=\aa\bb e^{z \am} - \frac {\sinh z\am}{z} \ap 
\label{ec}
\ee
are Casimir functions for this Poisson algebra.

The main consequence of the presence of the deformation in the
Poisson brackets (\ref{eb}) is that the phase space
realization (\ref{dd}) has to be deformed:
\bea
&&f_-^{(1)}=D(\am)=\ctea_1  q_1\qquad 
f_+^{(1)}=D(\ap)=p_1 e^{z {\ctea_1 q_1}}\cr
&& f_M^{(1)}= D(\bb)=\ctea_1 \qquad
f^{(1)}=D(\aa)=\frac {\sinh z {\ctea_1 q_1}}{z \ctea_1}  p_1  .
\label{ed}
\eea
Therefore, if we preserve the function ${\cal H}= \ap^2+
\am^2$, 
the one-particle Hamiltonian becomes a deformed oscillator:
\be
H^{(1)}_z=D({\cal H})= p_1^2 e^{2 z \ctea_1 q_1} +\ctea_1^2   q_1^2,
\label{ee}
\ee
and the phase space realization of the
Casimir is again zero $C^{(1)}_z=D({\cal C})=0$. In the spirit of this
method, (\ref{ee}) would be the basic non-standard ``integrable"
deformation of the harmonic oscillator linked to $U_z(h_4)$.

The $N=2$ integrable deformation is, as usual, a consequence of the  
two-particle phase space realization $D\otimes D$ acting on the
deformed coproduct of the generators. Namely,
\bea
&&f_-^{(2)}=\ctea_1   q_1+\ctea_2 q_2\qquad
f_M^{(2)}=\ctea_1  +\ctea_2  \cr
&&f_+^{(2)}=  p_1 \left( e^{z {\ctea_1 q_1}} + 2 z\ctea_2 
 \frac {\sinh z {\ctea_1 q_1}}{z \ctea_1} \right) e^{z {\ctea_2 q_2}}+
  p_2 e^{z {\ctea_2 q_2}}  e^{-z {\ctea_1 q_1}}\cr
&&f^{(2)}=\frac {\sinh z {\ctea_1 q_1}}{z \ctea_1} p_1 e^{z {\ctea_2 q_2}}
+\frac {\sinh z {\ctea_2 q_2}}{z \ctea_2} p_2 e^{-z {\ctea_1 q_1}} .
\label{ef}
\eea
Now, from ${\cal H}$, the deformed two-particle Hamiltonian reads:
\bea
&&H^{(2)}_z= 
 p_1^2 \left( e^{z {\ctea_1 q_1}} + 2 z\ctea_2 
 \frac {\sinh z {\ctea_1 q_1}}{z \ctea_1} \right)^2 e^{2 z
{\ctea_2 q_2}} +  p_2^2 e^{2 z {\ctea_2 q_2}}  e^{-2 z {\ctea_1 q_1}} 
+ \ctea_1^2 q_1^2+\ctea_2^2 q_2^2\cr 
&& 
\qquad  + 2\left\{ p_1 p_2 \left( e^{z {\ctea_1 q_1}} + 2 z\ctea_2 
 \frac {\sinh z {\ctea_1 q_1}}{z \ctea_1} \right)
 e^{2 z {\ctea_2 q_2}}  e^{-z {\ctea_1 q_1}} +
{\ctea_1 \ctea_2} {q_1 q_2} \right\} . 
\label{eg}
\eea
 A power series expansion of this Hamiltonian around
$z=0$ will give us an idea of the perturbation generated by the
non-standard deformation:
\be
H^{(2)}_z= H^{(2)} + 2z(p_1+p_2)\,\{(p_1+p_2)(\ctea_1 q_1 + \ctea_2 q_2)+
2\,q_1\,(\ctea_2 p_1 - \ctea_1 p_2)\}+o(z^2) 
\ee
where $H^{(2)}$ is the Hamiltonian (\ref{bg}).
A  constant of motion is given by the coproduct of the Casimir:
\be
C^{(2)}_z= (\ctea_1 p_2-\ctea_2 p_1)\left\{
\frac {\sinh z {\ctea_2 q_2}}{z \ctea_2} e^{z {\ctea_2 q_2}} - 
\frac {\sinh z{\ctea_1 q_1}}{z \ctea_1} e^{-z {\ctea_1 q_1}} \right\}.
\ee
Its power series expansion around $z=0$ is
\be
C^{(2)}_z=C^{(2)}+z(\ctea_1 p_2 - \ctea_2 p_1)(\ctea_1 q_1^2
+ \ctea_2 q_2^2)+ o(z^2) 
\ee
where $C^{(2)}$ is the non-deformed Casimir (\ref{dh}).

The $N$-dimensional result is now a matter of long but straightforward
computations. Firstly, we have to deduce the $m$-th coproduct of any
generator by applying the recurrence (\ref{fl}). From it, by
considering a $(D\otimes D\otimes\dots^{m)}\otimes D)$ representation, the
$m$-particle phase space realization of $U_z(h_4)$ is obtained in closed
form:
\bea &&f_-^{(m)}=
\sum_{i=1}^m \ctea_i q_i\qquad
f_+^{(m)}
=\sum_{i=1}^m  p_i \fun_i^{(m)}  e^{z \qq_i^{(m)}(q)}\cr
&&f_M^{(m)}=
\sum_{i=1}^m \ctea_i  \qquad
f^{(m)}
=\sum_{i=1}^m \frac {\sinh z {\ctea_i q_i}}{z \ctea_i} p_i e^{z
\qq_i^{(m)}(q)} 
\label{ei}
\eea
where the functions $\fun_i^{(m)}$ are defined as
\be
\fun_i^{(m)}= e^{z {\ctea_i q_i}} + 2 z 
\frac {\sinh z {\ctea_i q_i}}{z \ctea_i}
\sum_{l=i+1}^m \ctea_l .
\ee
From these expressions, we get the $N$-dimensional oscillator chain
induced by $U_z(h_4)$:
\be 
 H^{(N)}_z
=\sum_{i=1}^N \left( p_i^2 (\fun_i^{(N)})^2 e^{2z \qq_i^{(N)}(q)}
+\ctea_i^2   q_i^2  \right) + 2 \sum_{i<j}^N \left(p_i
p_j \fun_i^{(N)}\fun_j^{(N)} e^{ z \qq_{ij}^{(N)}(q)} +
\ctea_i\ctea_j  q_i q_j\right)
\label{em}
\ee 
whose integrals of motion are obtained from the coproduct of the Casimir.
They are
\bea
&& C^{(m)}_z=\sum_{i=1}^m p_i \frac {\sinh z{\ctea_i q_i}}{z \ctea_i}
\left\{ e^{ z {\ctea_i q_i}} e^{2z \qq_{+,i}^{(m)}(q)}
\sum_{j=1}^{i-1}\ctea_j +e^{- z {\ctea_i q_i}} e^{-2z
\qq_{-,i}^{(m)}(q)}\sum_{l=i+1}^{m}\ctea_l\right\}\cr &&\qquad -
\sum_{i=1}^m   p_i
 \frac{e^{2z \qq_{+,i}^{(m)}(q)} -  e^{-2z \qq_{-,i}^{(m)}(q)}}{2z} .
\label{en}
\eea

In conclusion, by using $h_4$ we have been able to derive the explicit
form for the constants of motion corresponding to the non-deformed
Heisenberg--Weyl Hamiltonian (\ref{bj}). This is not surprising,  since
$h_3$ is a sub-coalgebra of $h_4$, and the latter contains the full
information concerning the integrability of the system. However, the
situation is completely different at a deformed level, where quantum
$h_3$ algebras give by themselves integrable deformations of the
same system that differ from the ones that can be deduced by using
quantum $h_4$ algebras. This result can be understood if we recall that
$h_3$ quantum algebras cannot be obtained as sub-quantum $h_4$ algebras
(a comparison between the classification of quantum deformations for both
algebras \cite{Heis,Osc} clearly shows this fact). Although we have chosen
the closest representatives ($U_{z,\deforn}(h_3)$ and $U_z(h_4)$)
of both families, results here presented show that the integrable
deformations linked to both coalgebras lead to significantly distinct
structures. As a consequence, the obtention of all quantum algebra
deformations of a given $U(g)$ could be meaningful in the search for new
completely integrable systems.


\sect{Oscillator chains from $sl(2,\R)$ coalgebras}

It is well known that $sl(2,\R)$ can be considered as a dynamical algebra
for the one-dimensional harmonic oscillator. If $sl(2,\R)$ coalgebras and
their deformations are considered, a big class of new integrable
oscillator chains can be obtained. Among them, a set of systems
with anharmonic interactions arises in a very simple way.


\subsect{Non-deformed $U(sl(2,\R))$ coalgebra}

In the case of $sl(2,\R)$, we write the following non-deformed Poisson
coalgebra
\be \Delta(X)= X \otimes 1 +1\otimes X  
\qquad X\in \{\jj,\jm,\jp\} 
\label{fa}
\ee
\be
\{\jj,\jp\}=2\jp\qquad 
\{\jj,\jm\}=-2\jm\qquad \{\jm,\jp\}=4\jj 
\label{fb}
\ee
whose Casimir element is
\be
{\cal C}=\jj^2 - \jm \jp   .
\label{fc}
\ee

A phase space realization with vanishing Casimir is given by:
\be 
 f_-^{(1)}=D(\jm)=q_1^2\qquad 
f_+^{(1)}=D(\jp)= p_1^2\qquad
 f_3^{(1)}=D(\jj)=q_1 p_1 .
\label{fd}
\ee 
 From it, the harmonic oscillator Hamiltonian is
recovered if the following linear function of the generators of $sl(2,\R)$ 
\be {\cal H}= \jp+\para  \jm
\label{hsl}
\ee
is represented through (\ref{fd})
\be
H^{(1)}=D({\cal H})=   p_1^2 +\para q_1^2 .
\label{fe}
\ee

Now, the use of $D\otimes D$ onto the primitive coproduct (\ref{fa})
 leads to the two-particle phase space realization of $sl(2,\R)$
\be 
 f_-^{(2)}=q_1^2+q_2^2\qquad
 f_+^{(2)}= p_1^2 +  p_2^2\qquad
 f_3^{(2)}=q_1 p_1 + q_2 p_2 
\label{ff}
\ee
that, in turn, gives rise to the uncoupled oscillator Hamiltonian:
\be
H^{(2)}=  f_+^{(2)} +\para  f_-^{(2)} 
=  p_1^2+  p_2^2+\para (q_1^2+q_2^2).
\label{fg}
\ee
Note that now the frecuency of both oscillators is the same. The coproduct
of the Casimir will give us the corresponding integral of motion
\be
C^{(2)}=-({q_1}{p_2} -{q_2} {p_1})^2 
\label{fh}
\ee
that turns out to be the square of the angular momentum. (Note that, in
this particular case, we know that two more functionally independent
integrals exist, since the system (\ref{fg}) is known to be
superintegrable).

The construction of the $m$-dimensional particle phase space realization
is straightforward: \be
f_-^{(m)}=
\sum_{i=1}^m q_i^2\qquad
f_+^{(m)}
=\sum_{i=1}^m   p_i^2\qquad
f_3^{(m)}
=\sum_{i=1}^m q_i p_i .
\label{fi}
\ee
From it, the uncoupled chain of $N$ harmonic oscillators (all of them with
the same frecuency) is obtained:
\be 
 H^{(N)}
=    f_+^{(N)} +\para  f_-^{(N)} 
=\sum_{i=1}^N(  p_i^2 +\para q_i^2)
\label{fj}
\ee
together with the Casimirs $(m=2,\dots,N)$
\be
C^{(m)} =-\sum_{i<j}^m ({q_i}{p_j} - {q_j}{p_i})^2 .
\label{fk}
\ee


\subsect{Non-standard quantum deformation $U_z(sl(2,\R))$}

Now, we introduce a suitable Poisson realization of the non-standard
deformation of $sl(2,\R)$ \cite{Ohn} as follows:
\bea
&&\Delta(\jm)=  \jm \otimes 1+
1\otimes \jm \cr &&\Delta(\jp)=\jp \otimes e^{z \jm} + e^{-z \jm} \otimes
\jp \cr &&\Delta(\jj)=\jj \otimes e^{z \jm} + e^{-z \jm} \otimes \jj 
\label{ga}
\eea
\be 
\{\jj,\jp\}=2 \jp \cosh z\jm  \qquad 
 \{\jj,\jm\}=-2\frac {\sinh z\jm}{z}\qquad
\{\jm,\jp\}=4 \jj . 
\label{gb}
\ee 
The deformed Casimir reads
\be
{\cal C}_z=\jj^2  - \frac {\sinh z\jm}{z} \jp
\label{gc}
\ee
which vanishes under the
following one-particle deformed phase space realization of (\ref{gb}):
\be 
 f_-^{(1)}=D(\jm)=q_1^2\qquad 
f_+^{(1)}=D(\jp)=\frac {\sinh z q_1^2}{z q_1^2}  p_1^2  \qquad
f_3^{(1)}=D(\jj)=\frac {\sinh z q_1^2}{z q_1}  p_1 .
\label{gd}
\ee 

Let us now consider again the dynamical generator
${\cal H}= \jp +\para  \jm$. Under (\ref{gd}), we obtain a new deformed
oscillator
\be
H^{(1)}_z=D({\cal H})= \frac {\sinh z q_1^2}{z q_1^2} 
p_1^2  +\para q_1^2 .
\label{ge}
\ee

We follow the usual constructive method and we derive  the corresponding
two-particle phase space realization from the  
coproduct (\ref{ga}):
\bea
&&f_-^{(2)}=q_1^2+q_2^2\cr
&&f_+^{(2)}=
\frac {\sinh z q_1^2}{z q_1^2}  p_1^2  e^{z q_2^2} +
\frac {\sinh z q_2^2}{z q_2^2}  p_2^2  e^{-z q_1^2}\cr
&&f_3^{(2)}=
\frac {\sinh z q_1^2}{z q_1 }  p_1   e^{z q_2^2} +
\frac {\sinh z q_2^2}{z q_2 }  p_2  e^{-z q_1^2} .
\label{gf}
\eea
The associated two-particle Hamiltonian is
\be 
 H^{(2)}_z=  \frac {\sinh z q_1^2}{z q_1^2}  p_1^2  e^{z q_2^2} +
\frac {\sinh z q_2^2}{z q_2^2}  p_2^2  e^{-z q_1^2} +\para
  ( q_1^2+q_2^2  )
\label{gg}
\ee  
and the deformed coproduct for the Casimir leads to 
\be
  C^{(2)}_z=-\frac {\sinh z q_1^2 \sinh z q_2^2}{z^2 q_1^2 q_2^2}
\left({q_1}{p_2} - {q_2}{p_1}\right)^2 e^{-z q_1^2}
e^{z q_2^2}.
\ee

The $N$-dimensional generalization for this system is derived from the
 phase space realization of arbitrary dimension:
\be  
  f_-^{(m)}= \sum_{i=1}^m q_i^2\quad
 f_+^{(m)}=\sum_{i=1}^m
\frac {\sinh z q_i^2}{z q_i^2}  p_i^2  e^{z \kk_i^{(m)}(q^2) }\quad
f_3^{(m)}=\sum_{i=1}^m
\frac {\sinh z q_i^2}{z q_i}  p_i  e^{z \kk_i^{(m)}(q^2) }  
\label{gi}
\ee
where the $K$-functions we use in this section are defined in the appendix.

As a consequence, the $N$-dimensional Hamiltonian is just 
$ H^{(N)}_z =  f_+^{(N)}+\para f_-^{(N)}$
and the following constants of motion are deduced:
\be 
  C^{(m)}=- \sum_{i<j}^m 
\frac {\sinh z q_i^2 \sinh z q_j^2}{z^2 q_i^2 q_j^2}
\left({q_i}{p_j} - {q_j}{p_i}\right)^2  e^{ z
\kk_{ij}^{(m)}(q^2)} .
\label{gn}
\ee


\subsect{A class of integrable anharmonic chains}

Let us go back to the undeformed construction, and consider a
more general dynamical Hamiltonian ${\cal H}$ of the form
\be
{\cal H}=\jp+ {\cal F}(\jm),
\label{anh}
\ee
where ${\cal F}(\jm)$ is an arbitrary smooth function of $\jm$. The formalism
ensures that the corresponding system costructed from
(\ref{anh}) is also integrable, since ${\cal H}$ could be any function
of the coalgebra generators. Explicitly, this means that any
$N$-particle Hamiltonian of the form
\be 
 H^{(N)}
=    f_+^{(N)} +{\cal F}(f_-^{(N)}) 
=\sum_{i=1}^N p_i^2 + {\cal F}\left(\sum_{i=1}^N q_i^2\right)
\label{anha}
\ee
is completely integrable, being (\ref{fk}) its constants of motion.
Obviously, the linear case ${\cal F}(\jm)=\para\,\jm$ leads to the previous
harmonic case, and the quadratic one ${\cal F}(\jm)=\jm^2$ would give us an
interacting chain of quartic oscillators. Further definitions of the
function ${\cal F}$ would give us many other anharmonic chains, all of them
sharing the same dynamical symmetry and the same integrals of the motion.

Moreover, the corresponding integrable deformation of (\ref{anha}) is
provided by a realization of (\ref{anh}) in terms of (\ref{gi}):
\be 
 H_z^{(N)} 
=\sum_{i=1}^N
\frac {\sinh z q_i^2}{z q_i^2}  p_i^2  e^{z \kk_i^{(N)}(q^2) }
 + {\cal F}\left(\sum_{i=1}^N q_i^2\right)
\label{anhaz}
\ee
and (\ref{gn}) are again the associated integrals. This example shows
clearly the number of different systems that can be obtained
through the same coalgebra, and the need for a careful
inspection of known integrable systems in order to investigate their
possible coalgebra symmetries.


\subsect{Angular momentum chains and $sl(2,\R)$ coalgebras}

Finally, we would like to establish a connection between the
previous $sl(2,\R)$ oscillator chains and ``classical spin" systems. Such a
link is provided by the underlying coalgebra structure. If we substitute
the canonical realizations used until now in terms of angular
momentum realizations of the same abstract $sl(2,\R)$ Poisson coalgebra,
the very same construction will lead us to a long-range interacting
``spin chain" of the Gaudin type on which the
quantum deformation can be easily implemented.   

In particular, let us consider the realization $S$  
\be
g_3^{(1)}=S(\jj)=\s_3^1 \qquad
g_+^{(1)}=S(\jp)=\s_+^1 \qquad
g_-^{(1)}=S(\jm)=\s_-^1
\label{ha}
\ee
where the classical angular momentum variables $\sigma_i^1$ fulfil
\be
\{\sigma_3,\sigma_+\}=2\sigma_+\qquad 
\{\sigma_3,\sigma_-\}=-2\sigma_-\qquad \{\sigma_-,\sigma_+\}=4\sigma_3 
\label{pb}
\ee
and are
constrained by a given constant value of the Casimir function (\ref{fc}) in
the form $c_1=(\s_3^1)^2 - \s_-^1 \s_+^1$. 

As usual, $m$ different copies of
(\ref{ha}) (that, in principle, could have different values $c_i$ of the
Casimir)  can be distinguished with the aid of a superscript $\s_l^i$.
Then, the  $m$-th coproduct (\ref{fo}) provides the following realization
of the non-deformed $sl(2,\R)$ Poisson coalgebra :
\be
 g_l^{(m)}= (S\otimes \dots^{m)}\otimes
S)(\Delta^{(m)}(\s_l))=\sum_{i=1}^{m}{\s_l^i} 
\qquad l=+,-,3.
\label{hb}
\ee

Now, we can apply the usual construction and take ${\cal H}$ from
(\ref{hsl}). As a consequence, the uncoupled oscilator chain (\ref{fj}) is
equivalent to
\be 
H^{(N)} 
=  g_+^{(N)}+\para g_-^{(N)}=\sum_{i=1}^{m}{\s_+^i + \para \s_-^i} 
\label{hc}
\ee
and the Casimirs $C^{(m)}$ read $(m=2,\dots, N)$: 
\be 
 C^{(m)}= (g_3^{(m)})^2 - g_-^{(m)} g_+^{(m)} 
  = \sum_{i=1}^{m}{c_i} + 
 \sum_{i<j}^{m} (\s_3^i \s_3^j- \s_-^i \s_+^j -  \s_-^j \s_+^i ).
\label{hd}
\ee 
Note that these are just Gaudin Hamiltonians of the hyperbolic type
\cite{Gau,EEKT}.

A non-standard deformation of Gaudin system can be now obtained. The
deformed angular momentum realization corresponding to $U_z(sl(2,\R))$ is: 
\be
g_3^{(1)}=S(\jj)=\frac {\sinh z \s_-^1}{z \s_-^1}\s_3^1 \qquad
g_+^{(1)}=S(\jp)=\frac {\sinh z \s_-^1}{z \s_-^1} \s_+^1 \qquad
g_-^{(1)}=S(\jm)=\s_-^1
\label{ia}
\ee
where the classical coordinates $\s_l^1$ are defined on the cone  
$c_1=(\s_3^1)^2 -\s_-^1 \s_+^1=0$, that is, we are considering the zero
realization. 

It is easy to check that the $m$-th order of the coproduct (\ref{ga}) in
the above representation leads to the following functions
\bea  
 && g_-^{(m)}= \sum_{i=1}^m \s_-^i\qquad
 g_+^{(m)}=\sum_{i=1}^m
\frac {\sinh z \s_-^i}{z \s_-^i} \s_+^i  e^{z \kk_i^{(m)}(\s_-) }\cr
&&g_3^{(m)}=\sum_{i=1}^m
\frac {\sinh z \s_-^i}{z \s_-^i} \s_3^i  e^{z \kk_i^{(m)}(\s_-) }
\label{ib}
\eea
that define the non-standard deformation of (\ref{hb}). Therefore, any
function of these three objects (with $m=N$) can be taken as the deformed
Hamiltonian, that will Poisson-commute with the following ``non-standard
Gaudin Hamiltonians"
\bea
&& C^{(m)}_z=(g_3^{(m)})^2 - \frac {\sinh z g_-^{(m)}}{z} g_+^{(m)}\cr
&&\qquad=
\sum_{i=1}^m \left(\frac {\sinh z \s_-^i}{z \s_-^i}\right)^2
 e^{2 z \kk_i^{(m)}(\s_-) }\left\{ (\s_3^i)^2 -\s_-^i \s_+^i\right\}\cr
&&\quad\qquad +\sum_{i<j}^m \frac {\sinh z \s_-^i \sinh z \s_-^j}{z^2
\s_-^i \s_-^j}
 e^{  z \kk_{ij}^{(m)}(\s_-) } (\s_3^i \s_3^j - \s_-^i \s_+^j
- \s_+^i \s_-^j) .
\label{ic}
\eea
Since we are working in the zero representation with
$(\s_3^i)^2 -\s_-^i \s_+^i=0$, (\ref{ic}) can be simplified 
\be
 C^{(m)}_z=\sum_{i<j}^m \frac {\sinh z \s_-^i \sinh z \s_-^j}{z^2
\s_-^i
\s_-^j}
 e^{  z \kk_{ij}^{(m)}(\s_-) } (\s_3^i \s_3^j - \s_-^i \s_+^j .
- \s_+^i \s_-^j) .
\label{id}
\ee
These integrals are the angular momentum counterparts to
(\ref{gn}). Note that, as it was found for the standard case in \cite{BR},
the deformation can be interpreted as the introduction of a variable range
exchange in the model (compare (\ref{id}) with (\ref{hd})).


\bigskip
\bigskip

\noindent
{\Large{{\bf Acknowledgments}}}

\bigskip

\noindent The authors have been partially supported by DGICYT (Project 
PB94--1115) from the Ministerio de Educaci\'on y Ciencia de Espa\~na and
by Junta de Castilla y Le\'on (Projects CO1/396 and CO2/297) and
acknowledge discussions with Profs. O. Ragnisco, S. Chumakov
and G. Pogosyan.


\bigskip
\bigskip

\noindent
{\Large{{\bf Appendix}}}

\appendix

\setcounter{equation}{0}

\renewcommand{\theequation}{A.\arabic{equation}}

\bigskip

\noindent
The $\qq$-functions are defined as:
\bea
\qq_{-,i}^{(m)}(x)&=&\sum_{l=1}^{i-1} \ctea_l x_l\qquad
\qq_{+,i}^{(m)}(x)=\sum_{l=i+1}^m \ctea_l x_l
\label{zza} \\
\qq_i^{(m)}(x)&=&-\qq_{-,i}^{(m)}(x)+\qq_{+,i}^{(m)}(x)\cr
& =& - \sum_{k=1}^{i-1} \ctea_k x_k+ 
\sum_{l=i+1}^m \ctea_l x_l
\label{zzb}\\
 \qq_{ij}^{(m)}(x) & =&\qq_i^{(m)}(x)+ \qq_j^{(m)}(x)\cr
 & =& -(\ctea_i x_i - \ctea_j x_j)  - 2\sum_{k=1}^{i-1} \ctea_k x_k+
2\sum_{l=j+1}^m \ctea_l x_l  \qquad i<j .
\label{zzc} 
\eea

In the same way, the $\kk$-functions will be:
\bea
 \kk_i^{(m)}(x)& =&  - \sum_{k=1}^{i-1}  x_k+ 
\sum_{l=i+1}^m   x_l
\label{zzd}\\
 \kk_{ij}^{(m)}(x) & =&\kk_i^{(m)}(x)+ \kk_j^{(m)}(x)\cr
 & =& -(  x_i -   x_j)  - 2\sum_{k=1}^{i-1}   x_k+
2\sum_{l=j+1}^m   x_l  \qquad i<j .
\label{zze} 
\eea

Note that the following property is useful for computations involving
these functions
 \be \frac {\sinh(z\sum_{i=1}^m   x_i)}z = \sum_{i=1}^m  
\frac {\sinh z   x_i}z  e^{ z \kk_{i}^{(m)}(x)} .
\label{zzf}
\ee

\end{document}